\documentclass[12pt,letterpaper]{article}
\usepackage{booktabs,multirow}  
\usepackage{graphicx,lscape}
\usepackage{amsmath,amssymb}
\usepackage{mathtools}
\usepackage{bm}
\usepackage{threeparttable}
\usepackage{url}
\usepackage{graphicx}
\usepackage{soul}

\usepackage{color}

\usepackage{ulem} 

\usepackage{comment}

\newcommand{\blue}{\color{blue}}

\def\mybox#1{{\blue \vskip1mm \begin{center}
 \hspace{.0\textwidth}\vbox{\hrule\hbox{\vrule\kern6pt
  \parbox{.8\textwidth}{\kern6pt#1\vskip6pt}\kern6pt\vrule}\hrule}
  \end{center} \vskip-5mm}}

\begin{document}
\baselineskip=26pt

\begin{center}
\Large \bf On sample size determination for restricted mean survival time-based tests in randomized clinical trials
\end{center}

\vspace{3mm}

\begin{center}
Satoshi Hattori \\
{\it Department of Biomedical Statistics, Graduate School of Medicine and \\
Integrated Frontier Research for Medical Science Division, Institute for Open and Transdisciplinary ResearchInitiatives (OTRI), Osaka University \\
Yamadaoka 2-2, Suita City, Osaka 565-0871, Japan \\
E-mail:hattoris@biostat.med.osaka-u.ac.jp 
}\\
\vspace{6mm}
Hajime Uno \\
{\it 
Department of Medical Oncology, Division of Population Sciences, Dana-Farber Cancer Institute, \\
Department of Medicine, Harvard Medical School, and \\
Department of Data Science, Dana-Farber Cancer Institute, \\ Boston, Massachusetts 02215, the United States \\
E-mail: huno@ds.dfci.harvard.edu
}
\end{center}

\begin{center}
Running title: Sample size for RMST tests
\end{center}

\begin{center}
Version: 15Dec2022
\end{center}

\abstract{
Restricted mean survival time (RMST) is gaining attention as a measure to quantify the treatment effect on survival outcomes in randomized clinical trials. Several methods to determine sample size based on the RMST-based tests have been proposed. However, to the best of our knowledge, there is no discussion about the power and sample size regarding the augmented version of RMST-based tests, which utilize baseline covariates for a gain in estimation efficiency and in power for testing the no treatment effect.  
The conventional event-driven study design based on the log-rank test allows us to calculate the power for a given hazard ratio without specifying the survival functions. In contrast, the existing sample size determination methods for the RMST-based tests relies on the adequacy of the assumptions of the entire survival curves of two groups. Furthermore, to handle the augmented test, the correlation between the baseline covariates and the martingale residuals must be handled. To address these issues, we propose an approximated sample size formula for the augmented version of the RMST-based test, which does not require specifying the entire survival curve in the treatment group, and also a sample size recalculation approach to update the correlations between the baseline covariates and the martingale residuals with the blinded data. The proposed procedure will enable the studies to have the target power for a given RMST difference even when correct survival functions cannot be specified at the design stage.
}


\section{Introduction}
\label{sec1}
In randomized clinical trials designed to compare two treatments with a time-to-event outcome, the logrank test is extensively used for testing equality of the two event time distributions. To summarize the treatment effect magnitude, the hazard ratio (HR) is widely used, which is estimated with the Cox proportional hazards (PH) model\cite{Cox1972}. However, the PH assumption the Cox PH model requires is not necessarily satisfied in practice. Concerning the inappropriateness of the PH assumption, inference procedures for many kinds of non-proportional semiparametric models have been developed, including the accelerated failure time model \cite{Wei1992, Jin2003}, the proportional odds model \cite{Cheng1995} and the additive hazards model \cite{Lin1994}. These inference procedures were found to perform well in some practical situations. On the other hand, these semiparametric models also rely on some specific assumptions regarding the relationship between two event time distributions, such as additive hazards or proportional odds assumptions, and they are also subject to misspecification similar to the Cox PH model. They have also been rarely employed in confirmatory randomized clinical trials; instead the logrank/HR approach has been routinely used \cite{Uno2020}.

Many recent clinical trials of immunotherapies for cancer reported that the Kaplan-Meier curves of the two treatment arms are almost identical up to a certain time point from the randomization and after the time point, the two curves separated, indicating that the immunotherapy improved patients' survival \cite{Reck2016,  Guimaraes2020}. This late-onset efficacy reflects the mechanism of the immunotherapy; a certain duration is needed for the immunotherapy to act on the immune system. In this case, violation of the PH is essential because of the mechanism of the therapy. It motivates statisticians to consider more closely how to analyze the primary time-to-event endpoint in confirmatory randomized clinical trials under non-proportional hazards (non-PH). Uno et al. contrasted the pros and cons of several measures for the treatment effect alternative to the HR, including difference and ratio of the restricted mean survival time (RMST), which is the mean survival time truncated at a specific study time and calculated as the area under the survival curve from 0 to the truncation time point \cite{Uno2014, Uno2015, Royston2011, Royston2013, Tian2018}. 
Difference and ratio of RMSTs of the two groups can be good between-group contrast measures with clear clinical interpretation. These measures can be estimated non-parametrically without imposing any modeling assumptions, and thus are robust. 
This model-free property would be very attractive in confirmatory randomized clinical trials because the statistical analysis specified in the study protocol \cite{CPMP2003} would always give the intended interpretable results in the study protocol; the HR does not have this property once the PH assumption is violated. 
With the awareness with the issues of the HR, the RMST is gaining more attention in the clinical research community and is starting to be utilized in practice.  A study where the RMST-based analysis was used as the primary analysis can also be found\cite{Guimaraes2020} . 

There are several methods for calculating sample sizes available for the tests contrasting the RMST difference between two comparative groups.
\cite{Royston2013} showed a simulation-based method to determine the sample size for RMST. \cite{Uno2015} discussed a simulation-based method specifically for non-inferiority trials. 
\cite{Luo2019} and \cite{Eaton2020} discussed the use of an asymptotic power formula. 
These power calculation approaches require users to specify the entire survival curves of the two groups, 
where some simple parametric distribution have been assumed conventionally. 
For example, in \cite{Guimaraes2020}, the exponential distribution was assumed and the rate parameter of the exponential distribution was determined so that the 1 year survival rate became 0.855.  If another parametric model, for example, a log-normal model, were used for the power calculation, the resulting power would be different from the one based on the exponential distribution even if the 1 year survival rate was 0.855. 
For estimating the sample size more accurately, one may assume piecewise exponential distributions instead of simple parametric distributions as \cite{Luo2019} proposed. 
However, in practice, it would be still challenging to accurately specify the entire survival curves at the design stage due to limited information about the treatment.  If the specified survival curves are inaccurate, the sample size based on RMST-based tests may be under/over-estimated because the power formula \cite{Luo2019, Eaton2020} involves the entire survival curves of two groups.

On the other hand, the power formula for the log-rank test, or equivalent HR-based tests \cite{Schoenfeld1981} does not involve the entire survival curves but only a required number of events, anticipated HR, and type 1 error rate. 
Thus, when the log-rank test is used as the primary analysis, the final analysis is supposed to be conducted when the required number of events is observed to achieve the planned power. This is called an "event-driven study'' design \cite{Collett2004} and has been almost routinely used for decades. 
This approach still requires users to specify the entire survival curves of two groups, the anticipated accrual profile, and the follow-up duration to calculate the total number of subjects to enroll. Misspecification of the survival curves may lead to unexpected delays in the final analysis. 
However, because the final analysis is performed when the required number of events is observed and the power of the log-rank test depends on only the number of observed events, misspecification of the survival curves would not affect the power of the study. This is a practical advantage of using the log-rank test against RMST-based tests and weighted log-rank tests \cite{Yuan2020}. 

The potential usefulness of baseline covariates in the primary statistical analysis has been argued for a long time \cite{Tsiatis1985, DiRienzo2001, Pocock2002}. However, in most clinical trials, no or only a few baseline covariates are incorporated in the primary analysis; due to potential misspecification, regression analyses are hardly used and only limited number of covariates are adjusted with stratified analysis. The augmentation approach is gaining of much interest, in which covariates are used to reduce variations of estimators by attaching an augmented term to the estimators or the estimating equations \cite{Lu2008, Tsiatis2008, Zhang2008, Zhang2015, Tian2012, Jiang2019, Hattori2022}. With randomization, the addtion of the augmented term does not lead any bias to estimators and does not require any additional assumptions for validity. For the RMST-based tests, \cite{Tian2012} and \cite{Jiang2019} discussed the inference procedure. 
However, to the best of our knowledge, there has been no discussion about power and sample size calculations for the RMST-based tests with the augmentation. 

In this paper, we sought to develop a procedure that does not require a correct specification of survival functions at the design stage but can allow the study to achieve a target power for detecting a given RMST difference by monitoring data without unblinding. The proposed method shares the advantage of even-driven studies that do not require the correct specification of the survival functions from two groups. Similar to the event-driven studies, the proposed approach will inform the study investigators of the timing of the final analysis to achieve the target power. Regarding censoring, the proposed method assumes a common censoring distribution for two groups similar to the event-driven study with the logrank test.  To this end, we derive the asymptotic power formula under a local alternative.
With this formula, given the target type 1 error rate and the power, the sample size is determined by an anticipated difference in RMST, the survival distribution of the control group, censoring distribution, and a correlation among baseline covariates and the martingale residuals in addition to those parameters required for the unadjusted version. 
To estimate the correlation, one may utilize data from disease registries, which are recently available in many fields. 
We refer to such a dataset as the {\it reference data}, whereas the study we are designing is called the {\it target study}. This approach can be taken at the design stage. If the {\it reference data} well reflects the distribution of the baseline covariates and their associations with the time-to-event outcome in the {\it target study}, 
the power based on the proposed formula will be accurate for the {\it target study}. Otherwise, it will not be accurate.  Therefore, along with the power formula, we also propose a method to recalculate the sample size in the middle of the study with the blinded data. Since this sample size calculation is performed without breaking the blindness of the assigned treatment, the integrity of the study will be intact and the impact on the type I error rate will be negligible. 

The organization of the paper is as follows. 
Although our development covers the augmented RMST-based test, our consideration on sample size calculation would be useful even with the unadjusted RMST-based test. 
In Section 2.1, we begin with summarizing the asymptotic properties of the unadjusted test for the RMST difference, and in Section 2.2, introduce the augmented RMST-based test. 
In Section 3, we derive the asymptotic power formula under a local alternative. 
In Section 4, we demonstrate the sample size calculation with the power formula for the unadjusted test and the augmented tests in Section 4.1, followed by a mid-trial sample size modification procedure for the augmented test in Section 4.2. 
In Section 5, we report results of a simulation study. 
In Section 6, we demonstrate the application of the proposed methods to a real data. 
We conclude our paper by mentioning some limitations and the potential future direction of the research in Section 7. 
All the theoretical arguments are given in the Appendix.

\section{RMST-based tests}
\subsection{Unadjusted test for the RMST difference}
Suppose we are interested in designing a randomized clinical trial with a time-to-event endpoint. We call the clinical trial the {\it target study}. We consider a two group comparison and let $Z$ be a binary random variable with $P(Z=1)=\pi$, which is coded as 1 and 0 if a subject is allocated to the treatment and control groups, respectively. Let $T$ and $C$ be a failure time of interest and potential censoring time, respectively. The failure time $T$ should be right-censored by $C$ and then $X=\min{(T,C)}$ and $\Delta=I(T \le C)$ are observable. A vector of baseline covariates is denoted by $V$. From randomization, we assume
\begin{eqnarray*}
{\bf Condition \ 1:} \ \  V \perp Z,
\end{eqnarray*}
where for arbitrary random variables $A_1$ and $A_2$, $A_1 \perp A_2$ implies independence of $A_1$ and $A_2$. In addition, we assume the standard assumption in survival analysis;
\begin{eqnarray*}
{\bf Condition \ 2:} \ \  C \perp T |Z,
\end{eqnarray*}
where for arbitrary random variables $A_1, A_2$, and $A_3$, $A_1 \perp A_2|A_3$ implies that $A_1$ and $A_2$ are conditionally independent given $A_3$. 

We assume that $n$ subjects are enrolled in the study. Let $n$ i.i.d. copies of $(X, \Delta, Z, V^T)$ denoted by $(X_i, \Delta_i, Z_i, V_i^T), i=1, 2, ..., n$, where the subscript $i$ represents the $i$th subject. Let $S_z(t)=P(T \ge t|Z=z)$ be the survival function of the group $z=0, 1$. Denote the corresponding hazards and cumulative hazards function by $\lambda_z(t)$ and $\Lambda_z(t)$, respectively. The counting process and the at-risk process are denoted by $N_i(t)=I(X_i \le t, \Delta_i=1)$ and $Y_i(t)=I(X_i \ge t)$, respectively. 
The RMST over the interval $[0,\tau]$ for $Z=z$ is defined as $\theta_z=E\{min(T, \tau)|Z=z\}=\int_0^{\tau}S_z(t)dt$, where $\tau$ is a pre-specified truncation time. The RMST is estimated by $\hat{\theta}_z=\int_0^{\tau}\hat{S}_z(t)dt$, where $\hat{S}_z(t)$ is the Kaplan-Meier estimator for $S_z(t)$. To compare the two treatments, the RMST difference can be used, which is defined by $\theta=\theta_1-\theta_0$. It is estimated by $\hat{\theta}=\hat{\theta}_1-\hat{\theta}_0$.
As shown in Appendix A, the asymptotic variance of $\sqrt{n}(\hat{\theta}-\theta)$ is given by
\begin{align}
\sigma^2 &=
\int_0^{\tau} \frac{\{\int_t^{\tau}S_1(u)du\}^2}{E\{I(X \ge t)Z\}}  d\Lambda_1(t) 
+
\int_0^{\tau} \frac{\{\int_t^{\tau}S_0(u)du\}^2}{E\{I(X \ge t)(1-Z)\}}  d\Lambda_0(t). 
\nonumber \\
&=
\int_0^{\tau} \frac{\{\int_t^{\tau}S_1(u)du\}^2}{\pi S_1(t)G(t)}  d\Lambda_1(t) 
+
\int_0^{\tau} \frac{\{\int_t^{\tau}S_0(u)du\}^2}{(1-\pi)S_0(t)G(t)}  d\Lambda_0(t),
\label{sgm1} 
\end{align}
It is consistently estimated by 
\begin{align}
\hat{\sigma}^2_{1} &=
\int_0^{\tau} \frac{\{\int_t^{\tau}\hat{S}_1(u)du\}^2}{\bar{Y}_1(t)}  d\hat{\Lambda}_1(t) 
+
\int_0^{\tau} \frac{\{\int_t^{\tau}\hat{S}_0(u)du\}^2}{\bar{Y}_0(t)}  d\hat{\Lambda}_0(t),
\label{sgm2}
\end{align}
where $\bar{Y}_1(t)=n^{-1} \sum_{i=1}^n I(X_i \ge t)Z_i$, $\bar{Y}_0(t)=n^{-1} \sum_{i=1}^n I(X_i \ge t)(1-Z_i)$ and $\hat{\Lambda}_z(t)=\int_0^t \sum_{i=1}^n I(Z_i=z) \{n \bar{Y}_z(u)\}^{-1} dN_i(u)$ is the Nelson-Aalen estimates for $\Lambda_z(t)$.
Alternatively, $\sigma^2$ is also consistently estimated by $\hat{\sigma}^2_{2}=n^{-1} \sum_{i=1}^n \hat{H}_{i}^2$, where
\begin{eqnarray}
\hat{H}_{i} &=& - \int_0^{\tau} \frac{\int_t^{\tau} \hat{S}_1(u)du}{\bar{Y}_1(t)}Z_i d\hat{M}_{1,i}(t) 
+ \int_0^{\tau} \frac{\int_t^{\tau} \hat{S}_0(u)du}{\bar{Y}_0(t)}(1-Z_i) d\hat{M}_{0,i}(t),
\end{eqnarray}
where $\hat{M}_{z,i}(t)=N_i(t)-\int_0^t I(X_i \ge u) d\hat{\Lambda}_z(u)$ is the counting process martingale for $Z=z$. The derivation of $\hat{\sigma}^2_{1}$ and $\hat{\sigma}^2_{2}$ is given in Appendix A. 
We refer the test based on $\hat{\theta}$ as the unadjusted RMST test. 

\subsection{Augmented test for the RMST difference}

The augmented version of the RMST difference is defined as
\begin{eqnarray}
\hat{\theta}_{aug}(c) &=& \hat{\theta}-\frac{1}{n} \sum_{i=1}^n (Z_i-\pi) c^T V_i, \nonumber \\
&=& \hat{\theta}-AUG(c)
\label{rmst_aug}
\end{eqnarray}
where $c$ is a vector of the same dimension as $V$ (Tian et al. 2012). As argued in Section 2.1, the first term of ($\ref{rmst_aug}$) consistently estimates the true RMST difference $\theta$. For any $c$, the expectation of the second term is zero from Condition 1. Then ($\ref{rmst_aug}$) consistently estimates the true RMST difference for any fixed $c$. It is true even if $c$ is date-dependent as long as it converges to a constant in probability. We determine $c$ that minimizes the variance. The resulting minimizer is denoted by $\hat{c}$, which can be obtained by projecting the influence function of $\hat{\theta}$ onto the subspace of $L^2(dP)$ spanned by $\{(Z-\pi) V\}$, where $dP$ is the probability measure of the underlying probability space and $L^2(dP)$ is the Hilbert space of the all the square-integrable functions on the probability space. As argued in Appendix B, it is given by
\begin{eqnarray}
\hat{c}&=& \Big\{ \pi (1-\pi) \sum_{i=1}^n V_i V_i^T\Big\}^{-1} \nonumber \\
&\times& \sum_{i=1}^n (Z_i-\pi) V_i
\Big[
-Z_i\int_0^{\tau} \frac{\int_t^{\tau}\hat{S}_1(u)du}{\bar{Y}_1(t)} d\hat{M}_{1i}(t)
+(1-Z_i) \int_0^{\tau} \frac{\int_t^{\tau}\hat{S}_0(u)du}{\bar{Y}_0(t)} d\hat{M}_{0i}(t)
\Big]. \nonumber \\
\label{ccc}
\end{eqnarray}
Let $\hat{\theta}_{aug}=\hat{\theta}_{aug}(\hat{c})$. 
Then, $\theta$ is estimated by $\hat{\theta}_{aug}$ consistently and more efficiently than $\hat{\theta}$.  
The asymptotic variance of $\sqrt{n}(\hat{\theta}_{aug}-\theta)$ is consistently estimated by $\hat{\sigma}^2_{aug}=n^{-1} \sum_{i=1}^n \{\hat{H}_i-(Z_i-\pi) \hat{c}^T V_i\}^2$ (see Appendix B).

\section{Power formula for the unadjusted and augmented RMST tests}
To obtain a simple expression of the local power, we assume an additional condition, 
\begin{eqnarray*}
{\bf Condition \ 3:} \ \  C \perp Z.
\end{eqnarray*} 
This condition is assumed in the widely used power formula for the event-driven study by the logrank test \cite{Fleming1991}. 
Since $\hat{c}$ is derived by the orthogonal projection of the influence function, it holds that
\begin{eqnarray*}
\lim_{n \to \infty} Var(\sqrt{n}(\hat{\theta}_{aug}-\theta))
&=&
\lim_{n \to \infty} Var(\sqrt{n}(\hat{\theta}-\theta))
-
\lim_{n \to \infty} Var(\sqrt{n} AUG(\hat{c})) \nonumber \\
&=&Q_1-Q_2.
\end{eqnarray*}
Note that $Q_1$ agrees with $\sigma^2$ in ($\ref{sgm1}$). 
Suppose we are interested in testing the null hypothesis that the survival functions are common between the groups. It is denoted by $H_0: \log{\lambda_1(t)/\lambda_0(t)}=0$.
We consider the local alternative $H_1: \log{\lambda_1(t)/\lambda_0(t)}=\delta(t)/\sqrt{n}$, where  $\delta(t)$ is a deterministic function of time providing a specific alternative hypothesis of interest.

Under this alternative, $\sqrt{n}(\hat{\theta}-\theta_{alt})$ asymptotically has a zero-mean normal distribution with  variance $\sigma^2$ in the equation (\ref{sgm1}),
where
\begin{eqnarray*}
\theta_{alt} = \frac{\eta}{\sqrt{n}} =  \frac{1}{\sqrt{n}} \int_0^t \left\{ \int_0^v \delta(u)\lambda_0(u) du \right\}  S_0(v) dv.
\end{eqnarray*}

Under the local alternative and Conditions 2 and 3, it holds that $S_1(t)=S_0(t)+o(1)$, $\Lambda_1(t)=\Lambda_0(t)+o(1)$, $M_{1,i}(t)=M_{0,i}(t)+o_p(1)$, $E\{I(X \ge t)Z\}=S_1(t)G(t)\pi=S_0(t)G(t)\pi+o(1)$, and $E\{I(X \ge t)(1-Z)\}=S_0(t)G(t)(1-\pi)$. 
Applying these identities to ($\ref{sgm1}$), 
we can approximate $Q_1$ by 
$Q_1=\sigma^2=\tilde{\sigma_1}^2+o_p(1)$, where
\begin{align}
\tilde{\sigma}_{1}^2 &= \frac{1}{\pi(1-\pi)} \int_0^{\tau} \frac{\{\int_t^{\tau}S_0(u)du\}^2}{S_0(t)G(t)}d\Lambda_0(t). 
\label{tilde_sgm}
\end{align}
In Appendix C, we show that when $\pi=1/2$, it holds that 
\begin{eqnarray}
Q_2=\pi(1-\pi)e_2,
\label{q2}
\end{eqnarray}
where
\begin{eqnarray}
e_2&=& E\Big\{\int_0^{\tau} \frac{\int_t^{\tau}S_0(u)du }{S_0(t)G(t)} dM_0(t) V^T\Big\}
\{E(VV^T)\}^{-1} E\Big\{\int_0^{\tau}\frac{\int_t^{\tau}S_0(u)du }{S_0(t)G(t)} dM_0(t) V\Big\},
\nonumber \\
\label{e2}
\end{eqnarray}
and $M_0(t)=N(t)-\int_0^{\tau} Y(u)d\Lambda_0(u)$ is the martingale residuals under the null hypothesis. 
Thus, the variance of $\hat{\theta}_{aug}$ is asymptotically approximated by $v_{aug}^2=\{\tilde{\sigma}^2-\pi (1-\pi) e_2\}/n$. 
Then, the local power for a two-sided $\alpha$ level test is given by 
\begin{eqnarray}
\Phi(z_{\alpha/2}-\theta_{alt}/v_{aug})+1-\Phi(z_{(1-\alpha/2)}-\theta_{alt}/v_{aug}).
\label{power2}
\end{eqnarray}
For general $\pi$s other than 1/2, ($\ref{q2}$) with ($\ref{e2}$) is shown to hold with an additional assumption
\begin{eqnarray*}
{\bf Condition \ 4:} \ \  T \perp C| Z, V.
\end{eqnarray*}
A proof for the case of $\pi \ne 1/2$ is given in Appendix D. 

If one uses the unadjusted test, which is based on $\hat{\theta}$, the power is approximately calculated by setting $e_2=0$, or 
\begin{eqnarray}
\Phi(z_{\alpha/2}-\theta_{alt}/v)+1-\Phi(z_{(1-\alpha/2)}-\theta_{alt}/v),
\label{power3}
\end{eqnarray}
where $v^2=\tilde{\sigma}^2/n$.

Note that these power formulas, (\ref{power2}) and (\ref{power3}), are derived based on the local alternative hypothesis. These would provide good approximated power estimates only when the alternative is close to the null hypothesis.   However, this approximated approach is more convenient than the one using ${\sigma}^2$ for practice.  For example, for the power calculation of the unadjusted test, we need to specify the entire survival curves from the two groups ($S_1(t)$ and $S_0(t)$), and $G(t),$ when we use the approach based on ${\sigma}^2.$ On the other hand, the approximated approach using $\tilde{\sigma}_1^2$ requires us to specify only a between-group difference in RMST ($\eta$), $G(t),$ and the entire survival curve from the control group $(S_0(t)).$ This feature would be attractive to users since they would not have sufficient data to estimate the survival time distribution, especially in the treatment group at the study design stage.

\section{Sample size calculation}
\subsection{Sizing at the design stage}
In this subsection, we discuss sample size calculation for a randomized clinical trial with the RMST-based test at the design stage. We begin with the case in which the unadjusted RMST-based test for the primary analysis with two-tailed significance level of 0.05. We define the target sample size to maintain the target power $1-\beta$ for the minimum clinically meaningful difference $\theta_{alt}$. As given in ($\ref{power3}$), the power depends on the survival function of the control group $S_0(t)$ and the censoring distribution $G(t)$. 

To determine an accurate estimate of the sample size achieving the target power, one needs to carefully specify $S_0(t)$ and $G(t)$ with available information at the design stage. For example, the study team may only have datasets for the control group obtained from past studies or disease registries. We call it the {\it reference data}. 
Suppose we are planning a {\it target study} with the {\it reference data}.
When information for either the control or the experimental group is available, from the definition of $\tilde{\sigma}_{1}^2$ in ($\ref{tilde_sgm}$), one can approximately calculate the power by estimating $S_0(t)$ and $G(t)$ with the Kaplan-Meier method.

Next, we consider the case in which the augmented RMST test is used. To do so, we need to estimate $e_2$.
It depends on the martingale residuals under the null, which is free from the treatment allocation $Z$. Then, one can estimate $v_{aug}^2$ with a dataset of only the control group. Suppose we have $n_+$ subjects in the {\it reference data} and the same notation to Section 2 is used. Note that $Z=0$ for all the subjects. 
Then, the predicted power is obtained by replacing unknown quantities in ($\ref{e2}$). That is, it can be estimated by
\begin{eqnarray}
\hat{e}_2 &=&  \frac{1}{n_+} \sum_{i=1}^{n_+} \int_0^{\tau} \frac{\int_t^{\tau}\hat{S}_0(u)du }{\bar{Y}_0(t)} d\hat{M}_{0, i}(t) V_i^T
\Big\{\sum_{i=1}^{n_+}V_i V_i^T \Big\}^{-1} \sum_{i=1}^{n_+} \int_0^{\tau} \frac{\int_t^{\tau}\hat{S}_0(u)du }{\bar{Y}_0(t)} d\hat{M}_{0, i}(t) V_i. \nonumber
\end{eqnarray}

\subsection{Mid-trial sample size determination}
As demonstrated in Section 4.1, specification of the survival and censoring distributions $S_0(t)$ and $G(t)$ can be influential on the calculation of the predicted power for the unadjudted and then the augmented tests. Furthermore, as seen in the formula ($\ref{e2}$), the predicted power of the augmented test depends on the variance-covariance matrix of the covariates $V$ and the martingale residuals. Thus, in order that the sample size calculation by the predicted power accurately approximates the power of the {\it target study}, the {\it reference data} should be similar to the {\it target study} in these aspects. However, it is very hard to clarify that and find a definitely suitable {\it reference data}. With the notable feature that the power formula ($\ref{power2}$) and ($\ref{power3}$) are free from the treatment allocation $Z$, one can estimate the local power with the mid-trial blinded dataset. To be specific, we propose that at an early stage in the {\it target study} with $n_{mid}$ subjects ($n_{mid}<n$), $\tilde{\sigma}^2$ and $e_2$ are estimated only with available data, and calculated the predicted power for $n$ subjects. We can determine the sample size for the statistical analysis with the predicted power to be the target power, say 0.8. Since all the adaptations are made under a blinded review, it would avoid under or over-powered studies, maintaining integrity of the study with the nominal type 1 error rates. 

\section{Simulation study}
\subsection{Data generation}
We conducted a simulation study investigating the accuracy and effectiveness of the proposed power calculation methods. In this subsection, we explain how to generate three kinds of datasets ({\it sData 1-3}). We simulates a randomized clinical trial to compare two treatment groups with a time-to-event endpoint. 
 
Let $b_1$ and $b_2$ be independent random variables following the standard normal distribution. We generated two kinds of continuous covariates $V_1=b_1+\epsilon_1$ and $V_2=b_2+\epsilon_2$, where $\epsilon_1$ and $\epsilon_2$ followed the standard normal distribution independently. 
Independence among $b_1$, $b_2$, $\epsilon_1$ and $\epsilon_2$ was assumed. 

The {\it sData 1} and {\it sData 2} were generated from the marginal proportional hazard model under the null hypothesis of no treatment effect and the alternative hypothesis, respectively, as follows; the failure time $T$ was generated from the model,
\begin{eqnarray*}
\log{T}=\log{\{\lambda_0 (1-Z)+\lambda_1 Z\}}+\log{(-\log{U})},
\label{simu_t}
\end{eqnarray*}
where $Z$ was a binary random variable independent of $V_1$ and $V_2$, which represented the randomized treatment allocation with $P(Z=1)=1/2$. $U$ was a random variable, which might or might not be dependent on $V_1$ and $V_2$ and had the marginal uniform distribution on $(0,1)$. Thus the failure time distributions for $Z=1$ and $Z=0$ were the exponential distribution with the hazard $\lambda_1$ and $\lambda_0$, respectively. The hazard $\lambda_0$ is determined so that the corresponding 5 year survival rate was 0.2, and $\lambda_1$ is determined to satisfy the HR $\lambda_1/\lambda_0$ was 1 ({\it sData 1}) or 0.7 ({\it sData 2}). The random variable $U$ was generated under the following two settings; [V1] $U=\Phi_3 (b_1+b_2+\epsilon)$ and [V2] $U=\Phi_1(\epsilon)$, where $\epsilon$ is a standard normal random variable independent of $b_1$ and $b_2$ and $\Phi_m(.)$ is the cumulative distribution function of the zero-mean normal distribution with the variance of $m$. Then, under [V1] and [V2], the failure time $T$ satisfies the marginal Cox PH model. The potential censoring time $C$ was generated from the uniform distribution on $(0, 8)$. Under [V1], $T$ was dependent on covariates, whereas it was not under [V2]. The {\it sData 3} were generated under the non-PH. For $Z=0$, the failure time $T$ was generated from the same model as the {\it sData 1}; the exponential distribution of the hazard $\lambda_0$. For $Z=1$, $T$ was generated from the piecewise exponential distribution in a similar way to {\it sData 1}, in which  the hazard was $\lambda_0$ for $t<1$ and was $\lambda_2=-\log{(0.5)}/5$ for $1 \le t$. For each {\it sData}, 10,000 sets of 500 subjects were generated. 

The {\it sData 1} to  {\it sData 3} were regarded as the {\it target study}. We generated two kinds of {\it reference data} for each {\it sData}. One is from the same distribution of the control group, which is referred as {\it correctly-matched}.  The other was from a biased sampling from the control group; subjects of $V_1<1$ and $V_2<1$ were only sampled, which is referred as {\it mis-matched}. Correspondingly to the datasets for the {\it target study}, 10,000 sets of the {\it reference data} were generated, each of which included 200 subjects.  

Suppose we are interested in comparing the two groups using the RMST with $\tau=5$. The true RMST difference was 0, 0.514 and 0.595 for {\it sData 1} to {\it sData 3}, respectively. 

\subsection{Accuracy of the predicted power calculated with the reference data at the design stage}
In Table 1, empiricial powers of the unadjusted and augmented RMST tests of $n=500$ based on 10,000 simulation datasets are presented. Summaries of the predicted power calculated with the {\it correctly-matched} and {\it mis-matched} {\it reference data} are also demonstrated. 
The empirical sizes were very close to the nominal level of 5 percent and inclusion of the augmentation term did not lead to inflation of the type 1 error rates. The augmented test had certain gains in power for datasets in which the failure time had dependence of {\it $v_1$} and {\it $v_2$} ({\it sData 2a and 3a}). The average of the predicted power with the {\it correctly-matched reference data} (denoted by {\it cPP}) were close to the corresponding empirical power. On the other hand, those with the {\it mis-matched reference data} ({\it mPP}) were not necessarily close.   

\subsection{Validity of adaptive choice of sample size under a mid-trial blinded review of the target study}
As observed in Section 5.2, if the {\it reference data} do not reflect the distributional structure of the {\it target study}, the predicted power might not approximate the power of the {\it target study} accurately. To the simulation datasets, we applied the proposed mid-trial sample size evaluation procedure in Section 4.3. At a mid-trial blinded review with $n_{mid}=100$ or $=200$, we estimated $S_0(t)$ and $G(t)$ with the pooled data of the two treatment groups. Then, we calculated the predicted power with $n=n_{mid}, n_{mid}+10, n_{mid}+20, \cdots$ and decided the minimum sample size with the predicted power to be the target power 0.8 as the sample size for the final analysis. Empirical sizes and powers of the blinded adaptive sample size choice procedure were shown in Table 2 with $n_{mid}=100$ and $=200$, respectively. The results for {\it sData 1} indicated that the empirical sizes were close to the nominal level of 0.05 in all the scenarios. From the results for {\it sData 2} and {\it sData 3}, the empirical powers were very close to the target power 0.8. Overall, the proposed method successfully controlled the power. 

We also made a similar evaluation for the augmented RMST-based test with the covariates $V_1$ and $V_2$. We selected the final sample size of the predicted power 0.8 by the augmented test and the results are summarized in Table 2. It indicated that the empirical sizes were close to the nominal level and the empirical powers were also close to the target one.
Thus, these results suggested that the blinded adaptive sample size choice procedure successfully controlled the power maintaining the validity. 
Table 2 shows the distributions of the sample size selected  by the mid-trial blinded review. With augmentation, the number of sample size might be reduced substantially. Variations of the sample sizes were smaller with $n_{mid}=200$ than with $n_{mid}=100$.

\section{Examples}
\subsection{Colon data}
In this section, we illustrate our proposing method with a dataset from a randomized clinical trial to compare efficacy and safety of the three adjuvant therapies of levamisole alone, levamisole plus fluorouracil (5-FU) and no therapy (observational group) in  resected stage B and C colorectal carcinoma \cite{Laurie1989, Moertel1990}, which is available as the $colon$ dataset in the {\it R} package {\it SURVIVAL}. We pretend to conduct a randomized clinical trial to compare levamisole plus 5-FU and levamisole alone, which is the {\it target study}. We regard the dataset of the observational group in the {\it colon} dataset as the natural history dataset available when designing the {\it target study} and use it as the {\it reference data}.

Suppose we compare the overall survival between the levamisole plus 5-FU group and levamisole alone by using the unadjusted RMST-based test with the two-tailed significance level of 0.05. We define $\tau=1825$ (days) and set the minimum clinically important difference as 150 (days) with respect to the RMST-difference. After excluding subjects with missing values in the covariates listed in Section 6.2, the observational group of the {\it colon} data contained 305 subjects. We used this dataset as the {\it reference data}. Among the 305 subjects, 164 died. Estimating $S_0(t)$ and $G(t)$ in ($\ref{tilde_sgm}$) with the {\it reference data}, we evaluated the predicted powers with the formula ($\ref{power3}$). As presented in Table 3, with $n=490$, the predicted power was more than 0.8 to detect the RMST-difference $\theta_{alt}=150$ (days). 

To see how influential the specification of $S_0(t)$ and $G(t)$ is on the calculation of the predicted power, we calculated the predicted power with the exponential distributions for $S_0(t)$ and $G(t)$ with the same 5 year survival rates as those from the Kaplan-Meier estimates, respectively. The 5 year survival and censoring probabilities $S_0(1825)$ and $G(1825)$ were estimated as 0.520 and 0.965 with the Kaplan-Meier method. If we assume the exponential distributions for the survival and censoring distributions, the corresponding hazard parameters were $\lambda_{S}=3.58 \times 10^{-4}$ and $\lambda_{G}=1.95 \times 10^{-5}$, respectively. 
The predicted power based on these exponential survival and censoring distributions with $n=490$ was 0.759 to detect the RMST difference of 150 (days), suggesting that inappropriate specification of $S_0(t)$ and $G(t)$ can lead to over or under-estimation of the sample size. 

Based on the calculation with the {\it reference data}, we set $n=490$ as the target sample size. Concerning discrepancy between the {\it reference data} and the {\it target study}, we applied the mid-trial re-evaluation procedure following the method in Section 4.3. The predicted powers with the first 200 subjects were shown in Table 3. With $n=440$, the predicted power attained the target power 0.8. 

Next, we determined the target sample size using the augmented RMST-based test. \cite{Moertel1990} reported several prognosis factors in their Table 1 including $extent$ of local spread (submucosa, muscle, serosa, contiguous structures), the number of lymph {\it nodes} with detectable cancer, {\it differentiation} of tumor (well, moderate, poor), {\it obstruction} of colon by tumor, {\it perforation} of colon, {\it adherence} to nearby organs as well as {\it sex} and {\it age}. To determine the covariates included in the augmented term, we used a stepwise variable increase method. In the first stage, we considered the augmented RMST-based test with a single covariate and selected the covariate attaining the maximum $\hat{e}_2$ over all covariates. In the second stage, we selected the covariate providing the maximum gain in $\hat{e}_2$ by adding to the covariate selected in the first stage. This step was continued until all covariates were included in the model. In Table 4, the covariates selected in the process and the predicted power with $n=490$ are shown. In the first stage, {\it nodes} had the maximum gain in $\hat{e}_2$. In the second stage, we evaluated gains in $\hat{e}_2$ by adding one more covariate to {\it nodes}, and selected {\it differentiation}. Table 4 indicates that improvement in power was saturated at stage 3; the predicted power with the set of covariates of {\it nodes}, {\it differentiation} and {\it extent} was almost the same as that with all eight covariates. As seen in equation ($\ref{e2}$), the inverse of $E(VV^T)$ must be taken to calculate the predicted power. Thus, unnecessary variables should not be included in the augmented term to avoid a colinearity problem. In the present case, we selected three covariates. 

In Table 3, the predicted powers of the augmented test with the three covariates based on the {\it reference data} are shown; with $n=390$, the predicted power attained the target power of 0.8. The results of the mid-trial re-evaluation at $n_{mid}=200$ are also shown in Table 3, indicating that $n=410$ attains the target power of 0.8 and is suggested as the sample size with which
the final statistical analysis is conducted. In this example, the recommended sample size with the mid-trial blinded evaluation were not so different from that with the {\it reference data} for the augmented test. 
We compared the RMSTs of the two groups with $n=410$ subjects and summarized the results of the estimation in Table 5. The unadjusted estimate of the RMST difference was 130.2 (95 $\%$ confidence interval (CI): 15.6, 244.7) days. The augmented one with the selected three covariates gave the estimate 138.0 (95 $\%$ CI: 28.6, 247.3). The addition of the augmentation term made the length of the confidence interval certainly shorter. The augmented estimate with all eight covariates had a similar standard error to that with the selected three covariates.

\subsection{The Oak study}
The Poplar study is an open-label phase 2 study to compare the efficacy and safety of atezolizumab with docetaxel for non-small cell lung cancer \cite{Fehrenbacher2016}. Two hundred and eighty seven subjects were enrolled and were randomly assigned to one of the two treatments. The primary endpoint was overall survival and the HR of atezolizumab to docetaxel was estimated as 0.73 (95$\%$ CI: 0.53, 0.99; P=0.040). It was followed by the Oak study, which was a large-scale randomized confirmatory Phase 3 study to compare atezolizumab with docetaxel for non-small cell lung cancer \cite{Rittmeyer2017}. Eighty hundred and fifty patients were randomized to one of the two treatments. The HR of atezolizumab to docetaxel was estimated as 0.73 (95$\%$ confidence interval: 0.62, 0.87; P<0.001). As seen in Figure 3A for the Poplar study\cite{Fehrenbacher2016} and Figure 2A for the Oak study\cite{Rittmeyer2017}, there were delayed responses of immunotherapy observed and then the PH assumption did not seem to hold. We used these two studies\cite{Fehrenbacher2016,Rittmeyer2017} for illustrating our proposed methods.

The RMST difference with $\tau=18$ (months) was estimated as 1.22 (-0.23, 2.67) with P=0.099 in the Poplar study. 
We created an example study using the Oak study and the Poplar study. We regarded the Oak study as the {\it target study}, and the Poplar study as the {\it reference data}. Here, we created the {\it reference data} pooling subjects of the two treatment groups in the Poplar study. We set the RMST difference with $\tau=18$ (months) as the treatment contrast and set 1.5 months as the target RMST difference. We begin with the unadjusted RMST test. By estimating $S_0(t)$ and $G(t)$ with the {\it reference data}, the sample size attaining $1-\beta=0.9$ with a two-tailed 5 percent significance level was calculated as 710. We re-evaluated the sample size with the mid-trial blinded sample size re-estimation of the Oak study, in which randomly selected $n_{mid}=200$ subjects were used. With the estimates of  $S_0(t)$ and $G(t)$ at the mid-trial re-estimation, the sample size was calculated as 860.

Next, we considered the augmented test for the RMST difference as the primary analysis and evaluate the power. We used eight covariates for augmentation; the number of metastatic site ({\it metastasis}), age at baseline ({\it age}), smoking status (current, previous and never) ({\it smoke}), sex ({\it sex}), histology (Non-small cell lung cancer, Squamous cell cartinoma) ({\it histology}), race (White, Asian, others)  ({\it race}), ECOG performance status (0 or 1) ({\it egoggr}),  baseline sum of the longest diameters ({\it blSLD}) and the number of prior chemotherapies (1 or 2) ({\it priortrt}). We applied the stepwise variable increase method introduced in Section 6.1 to select the variables included in the augmented term. The history of the selection is presented in Table 6, in which the variable selected at each stage and the predicted power with n=710 are presented. The predicted power seemed to be saturated at the step 5. Then, we selected the five variables of {\it metastasis}, {\it age}, {\it smoke}, {\it sex} and {\it histology}. When we included these, the predicted power was 0.947. The number of subjects attaining the target power 0.9 was calculated as 580 with the five covariates. The augmentation could reduce the number of subjects substantially. We re-evaluated the predicted power at the blinded review with the $200$ subjects in the Oak study. The sample size assuring the target power 0.9 was calculated as 750.

The Poplar and the Oak studies shared many inclusion criteria. However, there was a substantial difference between the predicted power calculation at the design stage with the Poplar study data and the mid-trial blinded sample size re-estimation with the Oak study data. The latter only allowed us to enroll stage IIIB and IV patients. It might be influential on the association between covariates and the overall survival. We applied a regression model for the RMST difference with the inverse probability censoring weighted method \cite{Tian2014}. We observed that $histology$ was significantly associated with the overall survival in the Oak study, but not in the Poplar study. Such difference of prognosis between the studies might affect the predicted power calculation. In Table 7, we show the RMST differences estimated with $n=580$ or $n=750$ subjects. With $n=580$, significance was marginal and the mid-trial sample size re-calculation seemed to successfully adjust the sample size.

\section{Discussion}
In randomized clinical trials with a time-to-event endpoint, the logrank-HR approach is routinely used. An advantage of this strategy is applicability of the event-driven study design, where the final analysis is conducted when the number of observed events from the study reaches the target. This approach achieves the target power to detect a given HR, if the PH assumption is correct. Whether the survival functions of both groups are correctly specified or not does not affect the power of the final analysis by logrank test or HR-based tests. On the other hand, the power of the conventional RMST-based test may be under or over-estimated if the survival functions are not correctly specified at the design stage. This may be a challenge when it is used for confirmatory clinical trials \cite{Yuan2020}. 

In this paper, we used a local power formula for the RMST-based test and proposed a method to determine the timing of the final analysis, which resulted in the target power for detecting the target effect size. Our method is based on the idea of the blinded sample size calculation, which is one of the most accurate adaptive design techniques with minimal risk of violation of study integrity \cite{FDA2016, FDA2019}. The proposed method would eliminate a drawback of the conventional RMST-based design and might make it more feasible to design confirmatory studies with the RMST. 

We demonstrated two applications. In the first example of the colon data, the two chemotherapy groups were regarded as comparative groups of the {\it target study} and the {\it reference data} was artificially created from the observational arm of the same study. Since the {\it reference data} was one of the randomized arm in reality, distributions of covariates should be similar. Despite this, the predicted power with the {\it reference data} was not necessarily close to the predicted power with the mid-trial sample size re-calculation. The situation in the second example can occur frequently; Phase 3 studies are designed with results of Phase 2 studies with similar inclusion criteria. We observed a substantial difference between the predicted power at the design stage calculated with the Phase 2 study and the mid-trial blinded sample size re-calculation. These inconsistencies might happened due to inconsistencies of the associations between covariates and overall survival. Re-evaluation of the sample size with updated predicted powers at mid-trial blinded reviews is highly recommended to assure the target power for the target treatment effect.  

The key idea of the proposed method was to estimate the local power with blinded data, which was called the predicted power. 
Recently, \cite{Hattori2022} proposed a method to determine the number of subjects to conduct a testing hypothesis with the augmented version of the logrank test. The predicted power was monitored and the analysis was conducted at the date when the predicted power attained the target power. Since the predicted power of the unadjusted logrank test was determined by the number of the events, this approach is an extension of the event-driven design.  Thus, the predicted power-based approach could be a unified way to design randomized clinical trials with a time-to-event endpoint so that it has the target power for the target treatment effect at the final analysis.

In confirmatory clinical trials, interim analysis is widely used to consider early establishment of efficacy and early stopping of the study. For the RMST, \cite{Lu2021} discussed the interim analysis methodology with the RMST. Our current development is limited to the blinded consideration. Further research is warranted on using the proposed method in combination with unblinded interim analysis methodology.

\section*{Acknowlegements}
\label{s:acknow}
The first author's research was partly supported by Grant-in-Aid for Challenging Exploratory Research (16K12403) and for Scientific Research(16H06299, 18H03208) from the Ministry of Education, Science, Sports and Technology of Japan. 

\section*{Data availability statement}
We used a dataset available to the public.

\newpage
\clearpage

\begin{table}[]
\caption[]
	{\textit{Empirical powers of the unadjusted and augmented RMST-based tests and the average predicted powers calculated at the design stage with the reference data over 10,000 simulated datasets; {\it Power} means empirical powers, {\it cPP} and {\it mPP} are the predicted power with the correctly matched and incorrectly matched {\it reference data}, respectively. 
	}}
\begin{center}
\begin{tabular}{cccccccc}
\hline
dataset & status & dependence & true   & test      & power & cPP    & mPP   \\
\hline
sData1a & Null   & V1, V2     & 0       & augmented & 0.053 & NA    & NA    \\
        &        &            &         & unadjusted  & 0.054 & NA    & NA    \\
sData1b & Null   & None       & 0       & augmented & 0.054 & NA    & NA    \\
        &        &            &         & unadjusted  & 0.054 & NA    & NA    \\
sData2a & PH     & V1, V2     & 0.51424 & augmented & 0.925 & 0.940 & 0.903 \\
        &        &            &         & unadjusted  & 0.843 & 0.860 & 0.873 \\
sData2b & PH     & None       & 0.51424 & augmented & 0.840 & 0.863 & 0.862 \\
        &        &            &         & unadjusted  & 0.840 & 0.860  & 0.860  \\
sData3a & nonPH  & V1, V2     & 0.59506 & augmented & 0.967 & 0.963 & 0.965 \\
        &        &            &         & unadjusted  & 0.913 & 0.907 & 0.948 \\
sData3b & nonPH  & None       & 0.59506 & augmented & 0.910 & 0.910 & 0.941 \\
        &        &            &         & unadjusted  & 0.910 & 0.908 & 0.940  \\
\hline
\end{tabular}
\end{center}
\end{table}

\clearpage

\begin{landscape}
\begin{table}[]
\caption[]
	{\textit{Empirical powers of the unadjusted and augmented RMST-based tests conducted at the adaptively selected sample size with the predicted power with the earliest $n_{mid}$ subjects under the blind review and summary of sample sizes over 10,000 simulated datasets. 
	}}
\begin{center}
\begin{tabular}{ccccccccccccc}
\hline
       &            &        & \multicolumn{2}{c}{RMST difference} &            &         &       & \multicolumn{5}{c}{Adaptively selected   sample size} \\
$n_{mid}$ & test       & status & true             & target           & dependence & dataset & power & min    & nq1    & median    & q3   & max   \\
\hline
100    & unadjusted & Null   & 0                & 0.514            & v1, v2     & sData1a & 0.056 & 220       & 390      & 420       & 440     & 640      \\
       &            &        &                  &                  & None       & sData1b & 0.055 & 210       & 390      & 420       & 450     & 570      \\
       &            & PH     & 0.514            & 0.514            & v1, v2     & sData2a & 0.794 & 270       & 420      & 440       & 470     & 630      \\
       &            &        &                  &                  & None       & sData2b & 0.786 & 240       & 420      & 440       & 470     & 580      \\
       &            & nonPH  & 0.595            & 0.595            & v1, v2     & sData3a & 0.795 & 200       & 450      & 470       & 500     & 650      \\
       &            &        &                  &                  & None       & sData3b & 0.789 & 200       & 450      & 470       & 500     & 600      \\
       &            &        &                  &                  &            &         &       &           &          &           &         &          \\
       & augmented  & Null   & 0                & 0.514            & v1, v2     & sData1a & 0.055 & 150       & 280      & 310       & 340     & 470      \\
       &            &        &                  &                  & None       & sData1b & 0.058 & 190       & 380      & 410       & 440     & 550      \\
       &            & PH     & 0.514            & 0.514            & v1, v2     & sData2a & 0.790 & 180       & 310      & 330       & 360     & 480      \\
       &            &        &                  &                  & None       & sData2b & 0.776 & 220       & 410      & 430       & 460     & 560      \\
       &            & nonPH  & 0.595            & 0.595            & v1, v2     & sData3a & 0.791 & 150       & 330      & 360       & 380     & 490      \\
       &            &        &                  &                  & None       & sData3b & 0.784 & 190       & 440      & 460       & 490     & 580      \\
       &            &        &                  &                  &            &         &       &           &          &           &         &          \\
200    & unadjusted & Null   & 0                & 0.514            & v1, v2     & sData1a & 0.056 & 280       & 410      & 430       & 450     & 570      \\
       &            &        &                  &                  & None       & sData1b & 0.052 & 320       & 410      & 430       & 450     & 530      \\
       &            & PH     & 0.514            & 0.514            & v1, v2     & sData2a & 0.810 & 350       & 440      & 450       & 470     & 570      \\
       &            &        &                  &                  & None       & sData2b & 0.797 & 360       & 440      & 450       & 470     & 560      \\
       &            & nonPH  & 0.595            & 0.595            & v1, v2     & sData3a & 0.806 & 380       & 470      & 480       & 500     & 600      \\
       &            &        &                  &                  & None       & sData3b & 0.798 & 380       & 470      & 480       & 500     & 570      \\
       &            &        &                  &                  &            &         &       &           &          &           &         &          \\
       & augmented  & Null   & 0                & 0.514            & v1, v2     & sData1a & 0.054 & 210       & 300      & 320       & 340     & 430      \\
       &            &        &                  &                  & None       & sData1b & 0.055 & 310       & 410      & 430       & 450     & 530      \\
       &            & PH     & 0.514            & 0.514            & v1, v2     & sData2a & 0.806 & 230       & 330      & 340       & 360     & 440      \\
       &            &        &                  &                  & None       & sData2b & 0.793 & 350       & 430      & 450       & 470     & 550      \\
       &            & nonPH  & 0.595            & 0.595            & v1, v2     & sData3a & 0.806 & 270       & 350      & 370       & 390     & 470      \\
       &            &        &                  &                  & None       & sData3b & 0.797 & 380       & 460      & 480       & 500     & 570     \\
\hline
\end{tabular}
\end{center}
\end{table}
\end{landscape}

\clearpage

\newpage

\begin{table}[]
\caption[]
	{\textit{Predicted powers of the unadjusted and augmented RMST-based test with the four selected covariates using the observational group of the colon data as the reference data with two-tailed $5\%$ significance level and $n=490$ to deptect the true RMST-difference of 150 (days). 
	}}
\begin{center}
\begin{tabular}{ccccc}
\hline
     & \multicolumn{2}{c}{Unadjusted}                 & \multicolumn{2}{c}{Augmented}                  \\
n   & design stage & blind review & design stage & blind review \\
\hline
360 & 0.676        & 0.719        & 0.776        & 0.754        \\
370 & 0.688        & 0.730        & 0.787        & 0.766        \\
380 & 0.700        & 0.742        & 0.798        & 0.776        \\
390 & 0.711        & 0.752        & {\bf 0.808}        & 0.787        \\
400 & 0.722        & 0.763        & 0.818        & 0.797        \\
410 & 0.732        & 0.773        & 0.827        & {\bf 0.807}        \\
420 & 0.743        & 0.783        & 0.836        & 0.816        \\
430 & 0.752        & 0.792        & 0.844        & 0.825        \\
440 & 0.762        & {\bf 0.801}        & 0.852        & 0.833        \\
450 & 0.771        & 0.810        & 0.860        & 0.841        \\
460 & 0.780        & 0.818        & 0.867        & 0.849        \\
470 & 0.789        & 0.826        & 0.874        & 0.857        \\
480 & 0.797        & 0.834        & 0.881        & 0.864        \\
490 & {\bf 0.805}        & 0.842        & 0.887        & 0.871        \\
500 & 0.813        & 0.849        & 0.893        & 0.877                  \\
\hline
\end{tabular}
\end{center}
\end{table}

\clearpage

\newpage

\begin{table}[]
\caption[]
	{\textit{Predicted powers of the unadjusted and augmented RMST-based tests with the colon data as the {\it reference data}; $\#$ implies the number of covariates included in the augmented term, $Power$ is the predicted power with $n=490$, which has the power of 0.8 for the unadjusted RMST-based test. $Variables$ indicates the covariates of maximum gain in power by adding sequentially. For example, in the augmented logrank test with a  single covariate, {\it nodes} had the maximum value of $\hat{e}_2$ and {\it +differentiation} implies {\it differentiation} gave the maximum gain in the value of $\hat{e}_2$ by adding a single covariate to {\it nodes}. 
	}}
\begin{center}
\begin{tabular}{clcc}
\hline
Step & Variables  & $e_2$       & Power \\
\hline
0    &           &          & 0.805 \\
1    & {\it +differentiation}   & 228207.8 & 0.821 \\
2    & {\it  +nodes}    & 865255.6 & 0.867 \\
3    & {\it  +local}    & 1136876  & 0.886 \\
4    & {\it  +sex}   & 1159761  & 0.887 \\
5    & {\it  +obstruction} & 1164897  & 0.888 \\
6    & {\it  +perforation}   & 1166504  & 0.888 \\
7    & {\it  +age}      & 1166505  & 0.888 \\
8    & {\it  +adherence}   & 1166509  & 0.888 \\
\hline
\end{tabular}
\end{center}
\end{table}

\clearpage

\newpage

\begin{table}[]
\caption[]
	{\textit{Results of comparison of the RMST difference of the two treatments in the {\it colon data} with $n=410$ based on the unadjusted and augmented tests; 
"augmented (selected)" and "augmented (all)" imply the augmented test with the three selected covariates and that with all the eight covariates, respectively.
}}
\begin{center}
\begin{tabular}{lccc}
\hline
Method                                         & RMST difference (SE)   & 95\%CI & P-value \\
\hline
unadjusted                                     & 130.2(58.5) & (15.6, 244.7) & 0.026   \\
augmented (selected) & 138.0(55.8) & (28.6, 247.3) & 0.013   \\
augmented (all)          & 139.4(55.3) & (31.0, 247.8) & 0.012  \\
\hline
\end{tabular}
\end{center}
\end{table}

\clearpage
\newpage

\begin{table}[]
\caption[]
	{\textit{Predicted powers of the unadjusted and augmented RMST-based tests with the Poplar study data as the {\it reference data}; $\#$ implies the number of covariates included in the augmented term, $PP$ is the predicted power with $n=710$, which has the power of 0.8 for the unadjusted RMST-based test. $Variables$ indicates the covariates of maximum gain in power by adding sequentially. For example, in the augmented logrank test with a  single covariate, {\it nodes} had the maximum value of $\hat{e}_2$ and {\it +differentiation} implies {\it differentiation} gave the maximum gain in the value of $\hat{e}_2$ by adding a single covariate to {\it nodes}. 
	}}
\begin{center}
\begin{tabular}{clcc}
\hline
Step & Variable   & $e_2$      & Power \\
\hline
0    &            &         & 0.900   \\
1    & {\it +metastasis} & 15.259  & 0.907 \\
2    & {\it +age}        & 72.743  & 0.933 \\
3    & {\it +smoke}      & 91.191  & 0.940 \\
4    & {\it +sex}     & 103.727 & 0.945 \\
5    & {\it +histology}  & 109.419 & 0.947 \\
6    & {\it +race}       & 114.078 & 0.949 \\
7    & {\it +blSLD}      & 115.638 & 0.950 \\
8    & {\it +ecogger}    & 116.686 & 0.950 \\
9    & {\it +prioritrt}  & 116.686 & 0.950 \\
\hline
\end{tabular}
\end{center}
\end{table}

\clearpage
\newpage

\begin{table}[]
\caption[]
	{\textit{Results of comparison of the RMST difference of the two treatments in the {\it Oak data}  based on the unadjusted and augmented tests with $n=580$, which was determined with the {\it reference data} and with $n=750$, which was determined with the mid-trial sample size re-estimation.
}}
\begin{center}
\begin{tabular}{ccccc}
\hline
n   & Method     & RMST difference (SE) & 95\%CI          & pvalue \\
\hline
580 & unadjusted & 0.985 (0537)          & (-0.067, 2.038) & 0.066  \\
      & augmented  & 1.052 (0.511)         & (0.050, 2.053)  & 0.040  \\
750 & unadjusted & 1.107 (0.466)         & (0.193, 2.021)  & 0.018  \\
    & augmented  & 1.102 (0.448)         & (0.225, 1.979)  & 0.014  \\
\hline
\end{tabular}
\end{center}
\end{table}

\clearpage

\appendix

\section{Consistency of $\hat{\sigma}_{1}^2$ and $\hat{\sigma}_{2}^2$}
From the martingale representation of the Kaplan-Meier estimator, it holds that
\begin{eqnarray*}
\sqrt{n_z}\{\hat{S}_z(t)-S_z(t)\} &=& -S_z(t) \sqrt{n_z} n^{-1} \sum_{i=1}^n \int_0^t \frac{dM_{z,i}(u)}{\bar{Y}_z(u)}+o_p(1),
\end{eqnarray*}
where $\bar{Y}_1(u)=n^{-1}\sum_{i=1}^n I(X_i \ge u)Z_i$ and $\bar{Y}_0(u)=n^{-1}\sum_{i=1}^n I(X_i \ge u)(1-Z_i)$. 
With this representation, simple algebraic manipulation entails that
\begin{eqnarray}
\sqrt{n}(\hat{\theta}_1-\theta_1) &=& \frac{\sqrt{n}}{\sqrt{n_1}} \int_0^{\tau} \sqrt{n_1} \{\hat{S}_1(t)-S_1(t)\} ds
\nonumber \\
&\simeq&
-\frac{1}{\sqrt{n}} \sum_{i=1}^n Z_i \int_0^{\tau} \frac{\int_{u}^{\tau}S_1(s)ds}{E\{I(X_i \ge u)Z_i\}} dM_{1,i}(u).
\label{mgl1}
\end{eqnarray}
Similarly, 
\begin{eqnarray}
\sqrt{n}\{\hat{\theta}_0-\theta_0 \} &\simeq& 
-\frac{1}{\sqrt{n}} \sum_{i=1}^n (1-Z_i) \int_0^{\tau} \frac{\int_{u}^{\tau}S_0(s)ds}{E\{I(X_i \ge u)(1-Z_i)\}} dM_{0,i}(u).
\label{mgl0}
\end{eqnarray}
The standard moment calculus of the counting process martingale (Fleming and Harrington 1991) entails that the asymptotic variance is given by $\sigma_{dif, 1}^2$ and that $\hat{\sigma}_{dif, 1}^2$ consistently estimates it.

\section{Derivation of $\hat{\lowercase{c}}$ and consistency of $\hat{\sigma}_{aug}^2$}
From ($\ref{mgl1}$) and ($\ref{mgl0}$),
\begin{eqnarray*}
&& \sqrt{n}\{\hat{\theta}_{aug}(c)-\theta\} = \sqrt{n}\{\hat{\theta}-AUG(c)\} 
\\ &\simeq&
\frac{1}{\sqrt{n}} \sum_{i=1}^n \Big[ \int_0^{\tau} \Big\{
 \frac{-Z_i\int_{u}^{\tau}S_1(s)ds}{E\{I(X_i \ge u)Z_i\}} dM_{1,i}(u) 
+ \int_0^{\tau} \frac{(1-Z_i) \int_{u}^{\tau}S_0(s)ds}{E\{I(X_i \ge u)(1-Z_i)\}} dM_{0,i}(u)
\Big\} \\
& & - c^T (Z_i-\pi)V_i
\Big].
\end{eqnarray*}
Then, the variance of $\sqrt(n)(\hat{\theta}_{aug}(c)-\theta)$ converges to
\begin{eqnarray*}
E\Big[
 \int_0^{\tau} 
\Big\{
\frac{ -Z \int_{u}^{\tau}S_1(s)ds}{E\{I(X_i \ge u)Z_i\}} dM_{1}(u)
+\frac{ (1-Z) \int_{u}^{\tau}S_0(s)ds}{E\{I(X_i \ge u)(1-Z)\}} dM_{0}(u)
\Big\}
- c^T (Z-\pi)V
\Big]^2.
\end{eqnarray*}
A simple algebraic manipulation gives us the minimizer as
\begin{eqnarray*}
c^* &=& [E\{ (Z-\pi)^2VV^T\}]^{-1} \times \\
&& E\Big[(Z-\pi)
\Big\{
\frac{ -Z \int_{u}^{\tau}S_1(s)ds}{E\{I(X_i \ge u)Z_i\}} dM_{1}(u)
+\int_0^{\tau} \frac{ (1-Z) \int_{u}^{\tau}S_0(s)ds}{E\{I(X_i \ge u)(1-Z)\}} dM_{0}(u)
\Big\}\Big],
\end{eqnarray*}
which is consistently estimated by $\hat{c}$ from the standard law of large number. 
It holds that $n^{-\frac{1}{2}}\sum_{i=1}^n(Z_i-\pi)\hat{c}^T V_i=n^{-\frac{1}{2}}\sum_{i=1}^n(Z_i-\pi) c_*^T V_i+o_p(1)$. Then, by the standard central limit theorem, the asymptotic normality of $\sqrt{n}(\hat{\theta}_{aug}-\theta)$ and the consistency of $\hat{\sigma}_{aug}^2$ holds.

\section{Derivation of (7) when $\pi=1/2$}
From ($\ref{ccc}$), it holds that
\begin{eqnarray}
\sqrt{n} AUG_2&=&\hat{c}^T \frac{1}{\sqrt{n}} \sum_{i=1}^n (Z_i-\pi) V_i \\
&=& 
\frac{1}{n} \sum_{i=1}^n (Z_i-\pi) V_i \Big[
-Z_i \int_0^{\tau} \frac{\int_{u}^{\tau}S_1(s)ds}{E\{I(X_i \ge u)Z_i\}}   dM_{1,i}(u) \nonumber \\
&& +(1-Z_i) \int_0^{\tau} \frac{\int_{u}^{\tau}S_0(s)ds}{E\{I(X_i \ge u)(1-Z_i)\}}   dM_{0,i}(u)
\Big] \label{aug22} \\
&\times& \Big\{
\pi (1-\pi) \frac{1}{n} \sum_{i=1}^n V_i V_i^T
\Big\}^{-1}
\frac{1}{\sqrt{n}} \sum_{i=1}^n (Z_i-\pi)V_i.
\label{aug23}
\end{eqnarray}
By simple algebra, ($\ref{aug22}$) equals to
\begin{eqnarray}
&& -\frac{1}{n} \sum_{i=1}^n (Z_i-\pi)^2 V_i^T 
\int_0^{\tau}
\Big[
\frac{\int_{u}^{\tau}S_1(s)ds}{E\{I(X_i \ge u)Z_i\}}   dM_{1,i}(u) 
+
\frac{\int_{u}^{\tau}S_0(s)ds}{E\{I(X_i \ge u)(1-Z_i)\}}   dM_{0,i}(u)
\Big] \nonumber \\ 
&&-
\pi \frac{1}{n} \sum_{i=1}^n (Z_i-\pi) V_i^T 
\int_0^{\tau}
\frac{\int_{u}^{\tau}S_1(s)ds}{E\{I(X_i \ge u)Z_i\}}   dM_{1,i}(u) 
\label{t2} \\
&&+
(1-\pi)  \frac{1}{n} \sum_{i=1}^n (Z_i-\pi) V_i^T
\int_0^{\tau} 
\frac{\int_{u}^{\tau}S_0(s)ds}{E\{I(X_i \ge u)(1-Z_i)\}}   dM_{0,i}(u).
\label{t3} 
\end{eqnarray}
As argued in the subsection 3.2, $S_1(t)=S_0(t)+o(1)$, $\Lambda_1(t)=\Lambda_0(t)+o(1)$, $M_{1,i}(t)=M_{0,i}(t)+o_p(1)$, $E\{I(X \ge t)Z\}=S_1(t)G(t)\pi=S_0(t)G(t)\pi=o(1)$, and $E\{I(X \ge t)(1-Z)\}=S_0(t)G(t)(1-\pi)$. With these relationships, 
\begin{eqnarray*}
(\ref{t2}) &\simeq& 
- \frac{1}{n} \sum_{i=1}^n (Z_i-\pi) V_i^T 
\int_0^{\tau}
\frac{\int_{u}^{\tau}S_0(s)ds}{S_0(u) G(u)}   dM_{0,i}(u) 
\end{eqnarray*}
and 
\begin{eqnarray*}
(\ref{t3}) &\simeq& 
\frac{1}{n} \sum_{i=1}^n (Z_i-\pi) V_i^T
\int_0^{\tau} 
\frac{\int_{u}^{\tau}S_0(s)ds}{S_0(u) G(u)}   dM_{0,i}(u) 
\end{eqnarray*}
and thus ($\ref{t2}$) plus (\ref{t3}) is $o_p(1)$. Then, it holds
\begin{eqnarray}
 (\ref{aug22}) &\simeq& -\frac{1}{n} \sum_{i=1}^n (Z_i-\pi)^2 V_i^T 
\int_0^{\tau} 
\Big[\frac{1}{\pi(1-\pi)}
\frac{\int_{u}^{\tau}S_0(s)ds}{S_0(u) G(u)}   dM_{0,i}(u) 
\Big] \nonumber \\
&\simeq& 
-\frac{1}{\pi(1-\pi)}E\Big[
(Z-\pi)^2 
\int_0^{\tau}  
\frac{\int_{u}^{\tau}S_0(s)ds}{S_0(u) G(u)}   dM_{0}(u) V^T
\Big] \nonumber \\
&=& 
-E\Big[
\int_0^{\tau} \frac{\int_{u}^{\tau}S_0(s)ds}{S_0(u) G(u)}dM_{0}(u) V^T 
\Big]
\label{t11}
\end{eqnarray} 
where the last equality holds since $(Z_i-\pi)^2=1/4=\pi(1-\pi)$ algebraically when $\pi=1/2$. 
In $(\ref{aug23})$, $n^{-1} \sum_{i=1}^n V_i V_i^T \simeq E(VV^T)$ and from Condition 1
\begin{align*}
Var\Big(
\frac{1}{\sqrt{n}} \sum_{i=1}^n (Z_i-\pi)V_i
 \Big) \simeq \pi(1-\pi) E(VV^T).
\end{align*}
Then, $Var(\sqrt{n} AUG_2)$ asymptotically agree with ($\ref{q2}$). 

\section{Derivation of (7) when $\pi \ne 1/2$}
To obtain ($\ref{t11}$),  $(Z_i-\pi)^2=1/4=\pi(1-\pi)$ with $\pi=1/2$ is needed. Then, the equality of ($\ref{t11}$) is represented as
\begin{align*}
-\frac{1}{\pi(1-\pi)}E\Big[
(Z-\pi)^2 
\tilde{M} V^T
\Big] 
= 
-E\Big[
\tilde{M} V^T 
\Big],
\end{align*}
where $\tilde{M}=\int_0^{\tau}  
\frac{\int_{u}^{\tau}S_0(s)ds}{S_0(u) G(u)}  dM_{0}(u)$. 
Other parts of arguments hold with $\pi \ne 1/2$. Then, to show ($\ref{q2}$), it is enough to show
\begin{align}
E\Big[
(Z-\pi)^2 
\tilde{M} V^T
\Big]= 
\pi(1-\pi)E\Big[
\tilde{M} V^T 
\Big]+o_p(1).
\label{identity}
\end{align} 
Denote $\tilde{M}=\int_0^{\tau} h(u) dM_{0}(u)$ with $h(u)=\frac{\int_{u}^{\tau}S_0(s)ds}{S_0(u) G(u)}$. By simple algebra, 
\begin{align}
& E\Big\{ (Z-\pi)^2 \tilde{M} V^T\Big\}= E\Big[
V^T (Z-\pi)^2 \Big\{
N(\tau)-\int_0^{\tau} h(u) Y(u)\exp{\Big(\frac{\delta(u)}{\sqrt{n}}Z\Big)} d\Lambda_0(u) 
\Big\}
\Big] \nonumber \\
& =
E\Big[
V^T (Z-\pi)^2 \Big\{
N(\tau)-\int_0^{\tau}Y(u) d\Lambda(u| Z, V)
\Big\}
\Big] 
\label{t01} \\
& +
E\Big[
V^T (Z-\pi)^2 \Big\{
\int_0^{\tau}Y(u) d\Lambda(u| Z, V)-\exp{\Big(\frac{\delta(u)}{\sqrt{n}}Z\Big)} d\Lambda_0(u)
\Big\}
\Big] 
\label{t02}
\end{align}
where $\Lambda(u| Z, V)$ is the true conditional cumulative hazard function given $Z$ and $V$ under the local alternative $H_1$. As $n$ goes to infinity, $\exp{(\frac{\delta(t)}{\sqrt{n}}Z)} \Lambda_0(t)= \Lambda_0(t)+o(1)$ holds. Since $\Lambda(u| Z, V)$ leads to the marginal cumlative hazard function $\exp{(\frac{\delta(u)}{\sqrt{n}}Z)} \Lambda_0(t)$, $\Lambda(t| Z, V)$ is represented as $\Lambda(t| Z, V)=\Lambda(t|V)+o(1)$, where $\Lambda(t|V)$ is a cumulative hazard function conditional on $V$, which is free from $Z$. 
The term ($\ref{t01}$) is $E[\int_0^{\tau}V^T(Z-\pi)^2 \{dN(u)-Y(u) d\Lambda(u|Z, V)\}]=0$ since it is the expectation of a martingale integral with respect to a filtration generated by $\{V, Z, N(t), Y(t)\}$. As argued, the term ($\ref{t02}$) is asymptotically equivalent to
\begin{align}
&E\Big[
V^T (Z-\pi)^2 
\int_0^{\tau} Y(u) h(u) \{d\Lambda(u|V)-d\Lambda_0(u)\}
\Big] \nonumber \\
&=
E\Big[
V^T(Z-\pi)^2 
\int_0^{\tau} h(u) E\{I(T \ge u)I(C \ge u|Z, V)\} \{d\Lambda(u|V)-d\Lambda_0(u)\}
\Big] \nonumber \\
&=
E\Big[
V^T(Z-\pi)^2 
\int_0^{\tau} h(u) P(T \ge u|Z, V)P(C \ge u|Z, V) \{d\Lambda(u|V)-d\Lambda_0(u)\}
\Big], 
\label{t03}
\end{align}
where the last equality holds from Condition 4. As $n$ goes to infinity, $P(T \ge t|Z, V)$ should converge almost surely to a survival function free from $Z$, which is denoted by $S(t|V)$, since otherwise the marginal hazard function corresponding to $P(T \ge u|Z)=\int_0^t P(T \ge u|Z, V=x)dF_V(x)$ does not satisfy the local alternative $H_1$, where $F_V(x)$ is the cumulative distribution function of $V$. Note that $F_V(x)$ is free from $Z$ by Condition 1. From Conditions 1 and 3, $C \perp Z|V$ holds. Then, ($\ref{t03}$) becomes
\begin{align}
& E\Big[
V^T(Z-\pi)^2
\int_0^{\tau} h(u) S(u|V)P(C \ge u|V) \{d\Lambda(u|V)-d\Lambda_0(u)\}
\Big] \nonumber \\
&= \pi(1-\pi)E\Big[
V^T 
\int_0^{\tau} h(u) S(u|V)P(C \ge u|V) \{d\Lambda(u|V)-d\Lambda_0(u)\}
\Big] \nonumber \\
&=\pi(1-\pi)E\Big[
V^T 
\int_0^{\tau} Y(u) h(u) \{d\Lambda(u|Z, V)-\exp{\Big(\frac{\delta(u)}{\sqrt{n}}Z\Big)} d\Lambda_0(u)\}
\Big]+o(1) \nonumber \\
&=\pi(1-\pi)E\Big[
V^T 
\tilde{M}
\Big] 
\label{t12} \\
& -
\pi(1-\pi)E\Big[
\int_0^{\tau} V^T Y(u) h(u) \{dN(u)-d\Lambda(u|Z, V)\}
\Big] +o(1), 
\label{t13} 
\end{align}
where the first equality holds from Condition 1.
The term ($\ref{t13}$) is zero since it is the expectation of a martingale integral. The term ($\ref{t12}$) agrees with is the right hand side of ($\ref{identity}$), which completes the proof.

\end{document}